\documentclass[12pt]{article}
\usepackage{a4}
\newcommand{\beao}{\begin{eqnarray*}}
\newcommand{\eeao}{\end{eqnarray*}}
\newcommand{\be}{\begin{equation}}
\newcommand{\ee}{\end{equation}}
\newcommand{\bea}{\begin{eqnarray}}
\newcommand{\eea}{\end{eqnarray}}
\newcommand{\beq}{\begin{eqnarray}}
\newcommand{\eeq}{\end{eqnarray}}
\newcommand{\nn}{\nonumber}

\newcommand{\ep}{\epsilon}

\newcommand{\la}{\lambda}

\newcommand{\Ref}[1]{(\ref{#1})}

\newcommand\Tr{{~\rm Tr~}}

\usepackage{epsfig}
\begin{document}
\title{Long range chromomagnetic fields at high temperature}
\author{S.~Antropov $^a$, M.~Bordag $^{b,}$\thanks{E-mail:
Michael.Bordag@itp.uni-leipzig.de},
 V.~Demchik $^{a,}$\thanks{E-mail: vadimdi@yahoo.com}~ and~V.~Skalozub $^{a,}$\thanks{E-mail: skalozubv@daad-alumni.de}
 \\
 {\small\textit{ $^a$ Dnipropetrovsk National University, 49010 Dnipropetrovsk, Ukraine}} \\
 {\small\textit{ $^b$ University of Leipzig, Institute for Theoretical Physics,}}\\
 {\small\textit{ Postfach 100 920, 04009  Leipzig, Germany}} \\
 }
%
\maketitle
\begin{abstract}
The magnetic mass of  neutral gluons in Abelian chromomagnetic
field at high temperature is calculated in $SU(2)$ gluodynamics.
It is noted that such type fields are spontaneously generated at
high temperature.  The mass is computed either from the
Schwinger-Dyson equation accounting for the one-loop polarization
tensor or in Monte-Carlo simulations on a lattice. In latter case,
an average  magnetic flux penetrating a plaquette is measured for
a number of lattices. Both calculations are in agreement with each
other and result in zero magnetic mass. Some applications of the
results obtained are discussed.
\end{abstract}
\section{Introduction}
Investigation of deconfinement phase of QCD is a  topic of actual
interest in modern high energy physics. For instance, it is
related to confinement models involving monopole condensation in
the dual superconductor scenario. Further it is discussed as
lowering the phase transition temperature \cite{Cea:2007yv}. It is
also of interest for the QCD phase diagram, for quark-gluon plasma
and related topics like the state of the early universe.

As it was discovered recently, in the  non-Abelian gauge theories
at high temperature a spontaneous vacuum magnetization happens.
This has been determined  either by analytic quantum field theory
methods \cite{Starinets:1994vi}, \cite{Enqvist:1994rm},
\cite{Skalozub:1996ax}, \cite{Skalozub:1999bf} or in lattice
simulations \cite{Demchik:2008zz}. This phenomenon is analogous to
the spontaneous creation of Abelian chromomagnetic field $B =
const$ discovered at zero temperature $T = 0$ by Savvidy
\cite{Savvidy:1977as}. However, as it is well known, this vacuum
is unstable.
 The instability follows from  the tachyon mode $p^2_{0} = p^2_{||}
- g B$ presenting  in the gluon spectrum,
\begin{equation} \label{spectrum} p^2_{0} = p^2_{||} + (2 n + 1) g
B, ~n = - 1, 0, 1,... , \end{equation}
where $p_{||}$ is a momentum component along the field, $B$ is
field strength, $g$ is gauge coupling constant. The evolution of
the instability is  resulted in a condensate. Thus, at $T = 0$ the
Abelian constant chromomagnetic field  is completely screened.
Then interesting question arises: whether high temperature can
suppress the instability inherent to such a state?

As it is occurred, the situation  changes  at finite temperature
$T \not = 0$ and the spectrum stabilization happens due to either
a gluon magnetic mass \cite{Skalozub:1999bf} or so-called
$A_0$-condensate which is proportional to the Polyakov loop
\cite{Ebert:1996tj}. These are the extensions of the Savvidy model
to the finite temperature case. In this way a possibility of
spontaneous generation of the strong temperature-dependent and
stable color magnetic fields of order $g B \sim g^4 T^2$  is
realized. Further investigations of quarks and gluons  at this
background are of interest if the generated classical fields are
long range ones. In fact, this means that the field scale is
larger than Compton's wave length of a particle.

The most essential field characteristics at finite temperature are
the electric (Debye) and magnetic masses responsible for screening
of long range color electric and magnetic fields, respectively.
Debye's mass of gluons in the field presence has been calculated
already (see, for instance, \cite{Bordag:2006pr},
\cite{Bordag:2008wp}). It was derived as some function of
temperature and field strength. As concerns the magnetic mass, it
requires an additional consideration that is the topic of the
present investigation. The point is that in the field presence it
is natural to divide  gluons in two types - the charged and the
neutral gluons, - which have different magnetic masses at high
temperatures. For the former fields, the nonzero magnetic mass has
been determined in one-loop order \cite{Skalozub:1999bf}. For the
latter one this problem is not solved finally.  The zero value has
been obtained in one-loop approximation in
Refs.\cite{Strelchenko:2004 eg}, \cite{Bordag:2006pr}. However,
the role of higher order corrections remains not investigated,
yet.  Just this point need to be studied in more details. In
sections 2 we calculate the neutral gluon magnetic mass by means
of quantum field theory methods. In section 3 the magnetic mass of
Abelian chromomagnetic field is computed by using Monte-Carlo
simulations. In both computations a zero magnetic mass is
obtained. This, in particular, means that the spontaneously
created at high temperature Abelian chromomagnetic fields are long
range similarly to the case of usual $U(1)$ magnetic field. The
discussion and possible applications are given in section 4.
\section{Magnetic mass of neutral gluons }
In this section we calculate the magnetic mass of neutral gluons
by using analytic quantum field theory methods. In this approach,
the gluon magnetic mass can be determined in the imaginary time
formalism through the polarization tensor (PT),
\be m^2_{magn.}(s) = < s | \Pi(k,B,T)|s >_{k_4 = 0, k^2 \to 0}.
\ee
Here, $k_4 = 2 \pi N T, N = 0, \pm 1, \pm 2 ,...$ is a Matsubara
frequency, $k^2  $ is a momentum squared. The mean value is
calculated in the two states of polarization (denoted as $s = 1$
and $s = 2$ \cite{Bordag:2005br}, \cite{Bordag:2006pr}) transverse
to the gluon momentum $k_\mu$.

In  actual calculation, we consider $SU(2)$ gluodynamics and
assume that the Abelian chromomagnetic field is directed along the
third axis in coordinate and internal spaces. In the Landau gauge
it is described by the potential
\begin{equation} \label{field} A^a_\mu = \delta^{a 3} (0,0,B x^1,0), ~B =
const. \end{equation}
This is  solution to classical field equations without  source
terms. So, such field can  create spontaneously. For chosen  field
configuration it is convenient to decompose the potential as
neutral $ A_\mu = A_\mu^3 $ and charged  $ W_\mu^{\pm} =
\frac{1}{\sqrt{2}}(A^1_\mu  \pm A^2_\mu)$ gluons, where $a =
1,2,3$ is a color index. Just for the latter gluons the tree-level
spectrum is written in Eq.\Ref{spectrum}.

General strategy of calculation and the particular definitions of
quantities to calculate are the same as in
Ref.\cite{Bordag:2006pr}. In the neutral PT case, there are no
states unstable in tree approximation. The instability (and
related to it the imaginary part of PT) appears because of the
tachyon mode propagating inside loops. In tree-approximation,
there are two transverse with respect to the gluon momentum
$k_\mu$ states of polarization $s = 1$ and $s = 2$
\cite{Bordag:2005br}, \cite{Bordag:2006pr}. We introduce the
notations: $l_4 = k_4$ and $ h^2 = l_1^2 + l_2^2$ are transverse
with respect to external field $(B = B_3)$  momentum components
and $k^2 = h^2 + l_3^2,~ l^2 = l_3^2 + l_4^2$. To consider the
behavior of the static modes $l_4 = 0$ in perpendicular with
respect to the field direction we set $l_3 = 0$. Then the magnetic
mass of them is defined as the mean values of the PT in the
states  $s$ calculated in the limit of $h^2 \to 0$:
\be \label{mneutral} m^2_{magn.}(s) = < s | \Pi(k,B,T)|s >_{k_4 = 0,
h^2 \to 0}. \ee

The neutral PT is calculated in the form given in Eq.(38) of Ref.
\cite{Bordag:2006pr}:

\be\label{exp} \Pi_{\la\la'}(k)=\sum_{i=1}^{10} \ \Pi^{(i)}(k,B,T) \
T^{(i)}_{\la\la'} \ ,\ee
where  $T^{(i)}_{\la\la'}$  are the structures out of the momentum
$k_\mu$, medium velocity $u_\mu$ and $\delta_{\la \la'}$ and
 the form factors $\Pi^{(i)}(k)$ depend on the external momentum
$k_\mu$  through the variables $l^2$ and $h^2$ at zero temperature
and $h^2, k_4$ and $k_3$ at finite temperature.

Within the developed formalism, the polarization tensor becomes
the  expression of the type
\be \label{Pist}\Pi_{\la\la'}(k)=\int_0^\infty\int_0^\infty \ ds \ dt
~ M_{\la\la'}(p,k) \langle \Theta_T(s,t)\rangle, \ee
where in $M_{\la\la'}(p,k)$ we collected all factors appearing
from the vertexes and from the lines except for that  going into
 $\Theta_T(s,t)$  (Eq.(59) in Ref. \cite{Bordag:2006pr}):
\bea \label{ThetaT1} \langle\Theta (s, t)\rangle_T &=& \sum\limits_{N
= -\infty}^{ +\infty} \langle \Theta (s, t) \rangle \ \ \exp\left(-
\frac{N^2}{4 T^2 (s + t)} + i \frac{k_4 t N}{(s + t) T} \right)
\nn\\
&&= \sum\limits_{ N = -\infty}^{ +\infty}\Theta_T (s, t) .
\eea
We introduced the notation $\Theta_T(s,t)$ which is the basic
function appearing in all form factors, $N$ is  discrete energy variable.
The function $\langle \Theta (s, t) \rangle $ is given in  Eq.(47) of Ref.\cite{Bordag:2006pr}:
\bea \label{Theta}\langle\Theta\rangle= \frac{\exp
\left[-k\left(\frac{st}{s+t}\delta^{||}+\frac{ST}{S+T}\delta^{\perp}\right)
k \right]}{(4\pi)^2(s+t)\sinh(s+t)} \eea with $S=\tanh(s)$ and
$T=\tanh(t)$.

Here and below, as in Ref.\cite{Bordag:2006pr}, for simplicity we
set the field strength $g B = 1$. That means we measure all
quantities in units of $g B$. To return to the dimensionfull
variables one has to substitute $s \to g B s$, etc.

The factors $M^{(i)} $ giving contributions to the matrix elements
of $\Pi$ for the infrared limit of interest in Eq.\Ref{mneutral}
to the states $s = 1, s = 2$ are \cite{Bordag:2006pr}:
\bea\label{Mia}
M_2&=&4 \frac{1 - \cosh(q) \cosh(\xi)}{(\sinh(q))^2} - 2 + 8 \cosh(q)\cosh(\xi), \nn \\
M_3&=&- 2\cosh(2q) \frac{\xi\sinh(\xi)}{q\sinh(q)} - 2 + 6\cosh(\xi)\cosh(q), \nn \\
M_5&=& - 2 + 2 \cosh(q)\cosh(\xi), \eea
where $q = s + t, \xi = s - t = q ( 2 u - 1)$ and $s = q u, t =
q(1 - u)$. The contributions to the particular polarization states
$s$ are (Eq.(142) Ref.\cite{Bordag:2006pr}):
\bea\label{Pitr}\langle s=1| \Pi(k) |s=1\rangle &=& h^2 \Pi_2, \nn
\\
\langle s=2| \Pi(k) |s=2\rangle &=& h^2 \left( \Pi_3 + \Pi_5
\right). \eea
In this way, these matrix elements are the product of $h^2$ and
expressions which have a finite limit for $h^2=0$,
\be \Pi_i=\Pi^{(0)}_i+O(h^2). \label{2h2}\ee
The quantities $\Pi^{(0)}_i$ were calculated in
\cite{Bordag:2006pr} numerically. In fact, it is possible to
calculate them analytically in terms of simple zeta functions.

From the above expressions \Ref{Pist}-\Ref{Mia}, in leading order
for $T\to\infty$, which picks just the $N=0$-contribution, we note
\be \Pi^{(0)}_i
    =\frac{g^2}{(4\pi)^{3/2}}\frac{T}{\sqrt{g B}}
    \int\limits_0^1 d u \int\limits_0^{\infty}
    \frac{d q \sqrt{q}}{\sinh(q)}\  M_i(q,u)\,. \label{ffi}
\ee
In these expressions, the integration over $u$ can be carried out
explicitly,
\be \Pi^{(0)}_i
    =\frac{g^2}{(4\pi)^{3/2}}\frac{T}{\sqrt{g B}}
   \int\limits_0^{\infty}
    \frac{d q \sqrt{q}}{\sinh(q)}\  M_i(q)\,  \label{ffii}
\ee
with
\bea
M_2(q)&=&-2-\frac{4}{q}\coth(q)+\frac{4}{\sinh(q)^2}+\frac{4}{q} \sinh(2q),\nn\\
M_3(q)&=&-2-\frac{2}{q^2}\cosh(2q)\left(-1+q\coth(q)\right)+\frac{3}{q} \sinh(2q),\nn\\
M_5(q)&=&-2+\frac{1}{q}\sinh(2q), \label{2M}\eea
and the $q$-integrations remain.

These expressions are formally divergent for $q\to\infty$. This
divergence results from the tachyonic mode. Here we have to
remember that all above formulas are written in Euclidean
representation (basically, for technical reasons). In fact, we
have to start from the Minkowski space representation which can be
reached by an 'Anti'-Wick rotation, $q\to q e^{i\pi/2}$. In the
Minkowski space representation the parametric integrals are
convergent using the usual '$i\ep$'- prescription. But then, the
contribution from the tachyonic mode in the loop cannot be
Wick-rotated since, in momentum space, the corresponding pole is
on the 'wrong' side of the imaginary axis of the momentum $p_0$.
However, it can be 'Anti'-Wick rotated delivering a exponentially
fast converging integral. The remaining part can be Wick rotated
as usual.  In this way, if starting from the Euclidean
representation, the tachyonic part must be 'Anti'-Wick rotated
twice, $q\to q e^{i\pi}$. The remaining part can be kept as is.
The subdivision into tachyonic and remaining parts must be done
according to the behavior for $q\to\infty$. There is a freedom
left of redistribution power like contributions. It can be used to
avoid singularities in $q=0$.

We make the following subdivision into tachyonic parts (parts A),
\bea
{M_2^{\rm A}}{}&=&\left[\frac{4}{q}\,e^q \right]\sinh(q),\nn\\
{M_3^{\rm A}}{}&=&\left[\left(\frac{2(1-q)}{q^2}+\frac{3}{q}\right)\,e^q-\frac{2}{q^2}\right]\sinh(q),\nn\\
{M_5^{\rm A}}{}&=&\left[\frac{1}{q}\,e^q\right]\sinh(q),
\label{2tp}\eea
and remaining parts (parts B), $M_i^{\rm B}=M_i-M_i^{\rm A}$.

Being inserted into \Ref{ffii}, after the 'Anti'-Wick rotation,
the A-parts constitute simple integrals which can be done
immediately (for the moment we drop the common prefactors),
\bea
\Pi^{(0), \rm A}_2&=&4i\sqrt{\pi} ,\nn\\
\Pi^{(0), \rm A}_3&=&5i\sqrt{\pi},\nn\\
\Pi^{(0), \rm A}_5&=&i\sqrt{\pi}\,. \label{2tp1}\eea
The integrals in the B-parts are directly well convergent. Their
calculation is a bit more difficult, but after a number of
transformations all can be taken into a form to be found in
tables. The results are
\bea \Pi^{(0), \rm B}_2&=&
4\sqrt{\pi}+\frac{\left(3-3\sqrt{2}-8\pi+2\sqrt{2}\pi\right)\zeta(\frac{3}{2})}{2\sqrt{\pi}},\nn\\
\Pi^{(0), \rm B}_3&=&
5\sqrt{\pi}+\frac{\left(6-6\sqrt{2}-4\pi+2\sqrt{2}\pi\right)\zeta(\frac{3}{2})}{2\sqrt{\pi}},\nn\\
\Pi^{(0), \rm B}_5&=&
 \frac{1}{2}\sqrt{\pi}\left(2+\left(-4+\sqrt{2}\right)\zeta(\frac{3}{2})\right)\,.
\label{2tp2}\eea
Together with the A-parts we get finally
\bea \Pi^{(0)}_2&=&\frac{g^2}{(4\pi)^{3/2}}\frac{T}{\sqrt{g
B}}\left[4(1+i)\sqrt{\pi}+
\frac{\left(3-3\sqrt{2}-8\pi+2\sqrt{2}\pi\right)\zeta(\frac{3}{2})}{2\sqrt{\pi}}\right]
,\nn\\
\Pi^{(0)}_3&=&\frac{g^2}{(4\pi)^{3/2}}\frac{T}{\sqrt{g B}}\left[
5(1+i)\sqrt{\pi}+\frac{\left(6-6\sqrt{2}-4\pi+2\sqrt{2}\pi\right)\zeta(\frac{3}{2})}{2\sqrt{\pi}}\right]
,\nn\\
\Pi^{(0)}_5&=&\frac{g^2}{(4\pi)^{3/2}}\frac{T}{\sqrt{g B}}\left[
i\sqrt{\pi}+
 \frac{1}{2}\sqrt{\pi}\left(2+\left(-4+\sqrt{2}\right)\zeta(\frac{3}{2})\right)\right]\,.
\label{2tp3}\eea
The corresponding numerical values,
\bea
\Pi^{(0)}_2&=&\frac{g^2}{(4\pi)^{3/2}}\frac{T}{\sqrt{g B}}\left(-5.80+7.09 i\right),\nn\\
\Pi^{(0)}_3&=&\frac{g^2}{(4\pi)^{3/2}}\frac{T}{\sqrt{g B}}\left(1.04-8.9 i\right),\nn\\
\Pi^{(0)}_5&=&\frac{g^2}{(4\pi)^{3/2}}\frac{T}{\sqrt{g
B}}\left(-4.21+1.8 i\right)\,\eea
were obtained already in \cite{Bordag:2006pr} (however, with a
wrong sign of the imaginary parts). 
%
The above expressions have to be used in Eq.\Ref{Pitr} to obtain
final result. The sum of $\Pi_3 + \Pi_5 $ equals, $\Pi_3 + \Pi_5 =
\left[-3.17 - 7.09  i \right]$. The imaginary part is signaling
the instability of the state because of the tachyon mode, and the
real one is responsible for the screening of transverse  gluon
fields. The real and  imaginary parts are of the same order of
magnitude. This is similar to the case of Landau's damping at
finite temperature.

Let us turn to the real part and substitute it in the
Schwinger-Dyson equation
\be\label{SDe} D^{-1}(k^2) = k^2 - \Pi(k) \ee
for the neutral gluon Green function. We obtain for the mean
values
\bea\label{SDes1}\langle ~s=1~|D^{-1}(h^2)|~s=1 ~\rangle &=& h^2 -
Re
( \Pi_2) ~h^2 \nn \\
&=& h^2 \left( 1 +  5.8 \frac{T}{\sqrt{g B}} \right) \eea
and
\bea\label{SDes2}\langle~ s=2~|D^{-1}(h^2)|~s=2~ \rangle &=& h^2 -
Re
( \Pi_3 + \Pi_5) ~h^2 \nn \\
&=& h^2 \left( 1 + 7.09 \frac{T}{\sqrt{g B}} \right). \eea
These are the expressions of interest.

Two important conclusions follow from Eqs.\Ref{SDes1},\Ref{SDes2}.
First, for the transverse modes in the field presence,  there is
no fictitious pole similar to that of in the one-loop
approximation for zero external field background at finite
temperature \cite{Kalashnikov:1982sc}. The external field acts as
some kind resummation removing this singularity. Second, there is
no  magnetic screening mass in one-loop order. The transverse
components of the gluon field remain long range in this
approximation, as at zero external field
\cite{Kalashnikov:1982sc}.

Possible resolutions of the zero one-loop magnetic mass are
obvious: 1) the mass is generated in some kind resummation of
perturbation series (as this is well known at zero external field
case); 2) there are no magnetic mass for neutral gluons as in the
case of usual  magnetic fields. The problem requires
nonperturbation methods of computation.

\section{Magnetic mass on a lattice}
\qquad To solve the problem formulated in the end of the previous
section, we calculate the magnetic mass of the Abelian
chromomagnetic field by using Monte-Carlo (MC) simulations on a
lattice. For this purpose we, following Ref.\cite{DeGrand:1981yx},
investigate the behavior of the average magnetic flux penetrating
a lattice plaquette oriented perpendicular to the magnetic field
direction. To introduce the classical magnetic field \Ref{field}
on a lattice we apply the twisted boundary conditions discussed
below.

In the MC simulations, we use the hypercubic lattice $L_t\times
L_s^3$ with hypertorus geometry. The standard Wilson action of the
$SU(2)$ lattice gauge theory is
\begin{eqnarray}
S_W=\beta\sum_x\sum_{\mu<\nu}\left[1-\frac{1}{2}\Tr\left[{\bf
U}_\mu(x){\bf U}_\nu(x+a\hat{\mu}){\bf
U}^\dag_\mu(x+a\hat{\nu}){\bf
U}^\dag_\nu(x)\right]\right],\\ \nn \label{Action}
\end{eqnarray}
where $\beta=4/g^2$ is the lattice coupling constant, $g$ is a
bare gauge coupling, ${\bf U}_\mu(x)$ is the link variable located
on the link leaving the lattice site $x$ in the $\mu$-th
direction. The link variables ${\bf U}_\mu(x)$ are $SU(2)$
matrices decomposed in terms of the unity, $I$, and Pauli
$\sigma_j$, matrices in color space,
\begin{eqnarray} \label{SU2decomp}
U_\mu(x)=IU_\mu^0(x)+i\sigma_j U_\mu^j(x)=\left(\begin{array}{cc}
\:\:\:\, U_\mu^0(x)+i U_\mu^3(x) &
U_\mu^2(x)+i U_\mu^1(x)\\
-U_\mu^2(x)+i U_\mu^1(x) &
U_\mu^0(x)-i U_\mu^3(x)\end{array}\right).
\end{eqnarray}

Next let us incorporate the external Abelian magnetic field \Ref{field} into this
formalism. As in  Refs.\cite{Demchik:2008zz},
\cite{Demchik:2006qj}  we represent the  field in terms of external fluxes $\varphi$.
 The constant homogeneous external flux
$\varphi$ in the third spatial direction can be introduced
 by applying the following twisted boundary conditions
(t.b.c.) \cite{Demchik:2006qj}:
\begin{eqnarray}\label{22}
& &U_\mu(L_t,x_1,x_2,x_3)=U_\mu(0,x_1,x_2,x_3),\\\nonumber &
&U_\mu(x_0,L_s,x_2,x_3)=U_\mu(x_0,0,x_2,x_3),\\\nonumber &
&U_\mu(x_0,x_1,L_s,x_3)=e^{i\varphi}
U_\mu(x_0,x_1,0,x_3),\\\nonumber &
&U_\mu(x_0,x_1,x_2,L_s)=U_\mu(x_0,x_1,x_2,0).
\end{eqnarray}
These  give
\begin{eqnarray}\nonumber
&&U_\mu^0(x)= \cases{\begin{array}{cc}
  U_\mu^0(x)\cos(\varphi)-U_\mu^3(x)\sin(\varphi) & $for $x=(x_0,x_1,L_s,x_3) $ and $ \mu=2 \\
  \!\!\!\!\!\!\!\!\!\!\!\!\!\!\!\!\!\!\!\!\!\!\!\!\!\!\!\!\!\!\!\!\!\!\!\!\!\!\!\!\!\!\!\!\!\!\!\!\!\!\!\!\!\!\!\!U_\mu^0(x) &
  \!\!\!\!\!\!\!\!\!\!\!\!\!\!\!\!\!\!\!\!\!\!\!\!\!\!\!\!\!\!\!\!\!\!\!\!\!\!\!\!\!\!\!\!\!\!\!$for other links$\\
\end{array},
} \\ \nonumber &&U_\mu^3(x)= \cases{\begin{array}{cc}
  U_\mu^0(x)\sin(\varphi)+U_\mu^3(x)\cos(\varphi) & $for $x=(x_0,x_1,L_s,x_3) $ and $ \mu=2 \\
  \!\!\!\!\!\!\!\!\!\!\!\!\!\!\!\!\!\!\!\!\!\!\!\!\!\!\!\!\!\!\!\!\!\!\!\!\!\!\!\!\!\!\!\!\!\!\!\!\!\!\!\!\!\!\!\!U_\mu^0(x) &
  \!\!\!\!\!\!\!\!\!\!\!\!\!\!\!\!\!\!\!\!\!\!\!\!\!\!\!\!\!\!\!\!\!\!\!\!\!\!\!\!\!\!\!\!\!\!\!$for other links$ \\
\end{array}.
} \\ \nonumber
\end{eqnarray} \nonumber
The edge links in all directions are identified
as usual periodic boundary conditions except for the links in the
second spatial direction for which the additional phase $\varphi$
is added (Fig. 1). In the continuum limit, such t.b.c. settle the
magnetic field with the potential $\bar{A}_\mu = (0,0, B x^1,0)$ \Ref{field}.
The magnetic flux $\varphi$ is measured in angular units and can
take  continuous values from $0$ to $2\pi$.

More details on the t.b.c. can be found in
Ref.\cite{DeGrand:1981yx}. In this  paper, the twist of the
boundary conditions was applied to introduce the magnetic flux of
the Dirac monopole. Then the  magnetic mass of this non-Abelian
magnetic field is measured by investigating the average plaquette
values for the twisted and untwisted lattices. The main object of
such type  investigations is the difference (magnetic flux through
a lattice plaquette perpendicular to the $OZ$ axis):
\be \label{uflux} \langle U_{untwisted}\rangle  - \langle
U_{twisted}\rangle = f(m, L_s), \ee
which is fitted for each lattice geometry $L_t\times L_s^3$ by
different functions $f(m, L_s)$. Below we follow this approach and
measure the magnetic mass of the Abelian field of interest.
\begin{figure}
\begin{center}
\includegraphics[bb=58 552 317 778,width=0.35\textwidth]{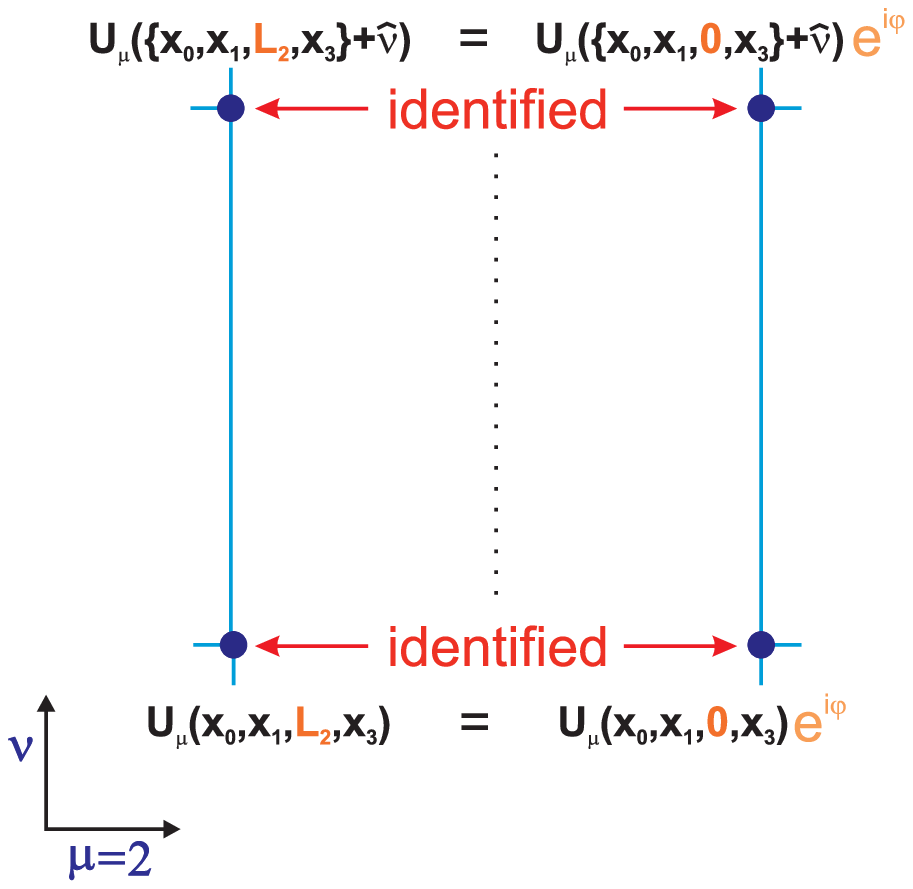}
\end{center}\label{Fig:1}
{\hskip 0.5cm {\bf Figure 1:} The plaquette presentation of the twisted boundary
conditions.}
\end{figure}
The temperature is introduced in a standard way through a lattice
asymmetry in the temporal direction ($L_t<L_s$). The measurements
were fulfilled for the value of $\beta = 2.6$ in the perturbation
regime for the deconfinement phase. Lattices with $L_t=4$ and
$L_s$ up to $32$ were used. To update the lattice, heat-bath
algorithm with overrelaxation was used \cite{Creutz:1987xi}. To
thermalize the system, up to $6000$ MC iterations were used. The
plaquette average is calculated by averaging up to $10000$ working
iterations.

To estimate the behavior of magnetic fields a large amount of
simulation data must be prepared. Unfortunately, traditional
computational resources are lack to perform the detailed analysis.
In our case, we use the General Purpose computation on Graphics
Processing Units (GPGPU) technology allowing to study large
lattices on personal computers. GPU programming model implemented
here and some technical details on MC simulations on ATI graphics
processing units (GPU) are given in Ref.~\cite{Demchik:2009ni}.

For GPU simulations, we apply GPU cluster of AMD/ATI Radeon GPUs:
HD4850, HD4870, HD5850 and HD5870. A peak performance of the GPU
cluster used is up to 8 Tflops. The common checkerboard scheme is
used for internal GPU parallelization. Simple parallelization
scheme is implemented on the cluster level -- each node in the
cluster performed independent MC simulation for parameters given
by the host system. Simulated data set is collected after each MC
run by the host system, as the host system we use one of cluster
nodes. Such scheme allows to increase linearly the performance of
cluster with increasing the number of nodes in the cluster.

To design the GPU-applications it must be accounted for that each
general-purpose register (GPR) and memory cell has the four 32-bit
components (single precision)  called GPR-slots and usually
designated as .x, .y, .z and .w. Thus, for $SU(2)$ model it is
natural to store all the components of link matrices
($U^0_\mu(x)$, $U^1_\mu(x)$, $U^2_\mu(x)$ and $U^3_\mu(x)$) (see
Eq.\Ref{SU2decomp}) as one GPU-cell. For example, if GPR $R1={\bf
U}_\mu(x)$, then the components of this register are
$R1.x=U^0_\mu(x)$, $R1.y=U^1_\mu(x)$, $R1.z=U^2_\mu(x)$ and
$R1.w=U^3_\mu(x)$. So, lattice data are stored with the single
precision, MC updating is performed with the single precision
whereas all averaging measurements were performed with the double
precision to avoid error accumulation.

Distinguishing feature of the employed program model is that all
data necessary for simulations are stored in GPU memory. GPU
carries out intermediate actions and returns the results to the
host program for final data handling and output. We avoid any data
transfer during the run-time between the host program and kernels
to speed-up the execution process.

To generate the pseudo-random numbers for MC procedure, three
different pseudo-random number generators are used: {\sf RANMAR},
{\sf RANLUX} and {\sf XOR128} \cite{Demchik:2010fd}. The last one
allows to obtain the maximal performance but is not widely used in
MC simulations. So, all the results were checked with the slower
generators {\sf RANMAR} and {\sf RANLUX}.

Performance analysis ~indicates ~that the GPU-based MC simulation
program shows better speed-up factors for big lattices in comparing
with the CPU-based one. For the majority  lattice geometries the GPU vs. CPU
(single-thread CPU execution) speed-up factor is above 50x and for
some lattice sizes could  overcome the factor 100x.

Thus, GPU-based MC program allows to calculate the difference
(\ref{uflux}) for a wide interval of lattice geometries. Also, up
to 1000 independent runs  for each lattice size were performed in
order to decrease the dispersion of the obtained values
$f(m,L_s)$. The whole set of simulation data for different lattice
geometries was fitted with the several functions which correspond
to the different behavior of magnetic flux.

The flux value $f(m, L_s)$ is determined by (\ref{uflux}) and
shown in Table 1 and Figure 2. The whole set of the data obtained
in MC simulations is divided into 15 bins. The mean values are
presented as the black points and the corresponding $2\sigma$
confidential level intervals are depicted by the vertical lines.
The plotted data refer to $L_t = 4$ and $\beta = 2.6$. The
relative errors decrease from $11.1\%$ to $4.1\%$ with increasing
$L_s$, as it should be.

In order to investigate different hypothesis for the behavior of
the magnetic flux we try to fit the MC data with some set of
functions by means of the $\chi^2$-method. The first test function
$C/r^2$ corresponds to the magnetic flux tube formation. Here $r$
is the lattice size $L_s$ in the $X$ and $Y$ directions, $C$ is an
unknown parameter. The total magnetic flux through the lattice is
conserved in this case. Next function $C/r^4$ describes the
Coulomb-like behavior and the function $C/r^2 \exp(- m^2 r^2)$ is
signaling the generation of the magnetic mass $m$
\cite{DeGrand:1981yx}. The last functions $C/r \exp(- m r)$ and
$C/r$ can be related to increasing of the field strength with a
temperature growth. This is because the total magnetic flux
through the lattice is growing faster than in the case of the
magnetic flux tube formation.

The fitted curves are shown in Figure 2. These are: a) the
visually coinciding solid curves -- $C/r$, $C/r \exp(-m r)$, $C/r
\exp(-m r^2)$; b) the dot curve -- $C \exp(-m r^2)$; c) the dash
curve -- $C/r^2$; d) the dash-dot curve -- $C/r^4$.

The numerical results of fitting procedure are collected in Table
2. The table contains the test functions, the values of the
$\chi^2$-function corresponding to the $95\%$ confidence level,
the obtained magnetic masses $m$ and parameters $C$.

As it follows from  Table 2, the best fit function is $C/r
\exp(-mr)$ with a small value of the magnetic mass $m=1.25\times
10^{-6}$. The value of $\chi^2$ function in this case is very
close to the $m = 0$ situation and statistically these cases are
indistinguishable. Really, the statistical errors are larger than
the fitted value of $m$. Thus, from the performed analysis we can
conclude that the neutral component of the gluon field is not
screened at high temperature like usual magnetic field. This
result is in agreement with that of previous section.

\begin{center}
\includegraphics[bb=89 497 496 678,width=0.85\textwidth]{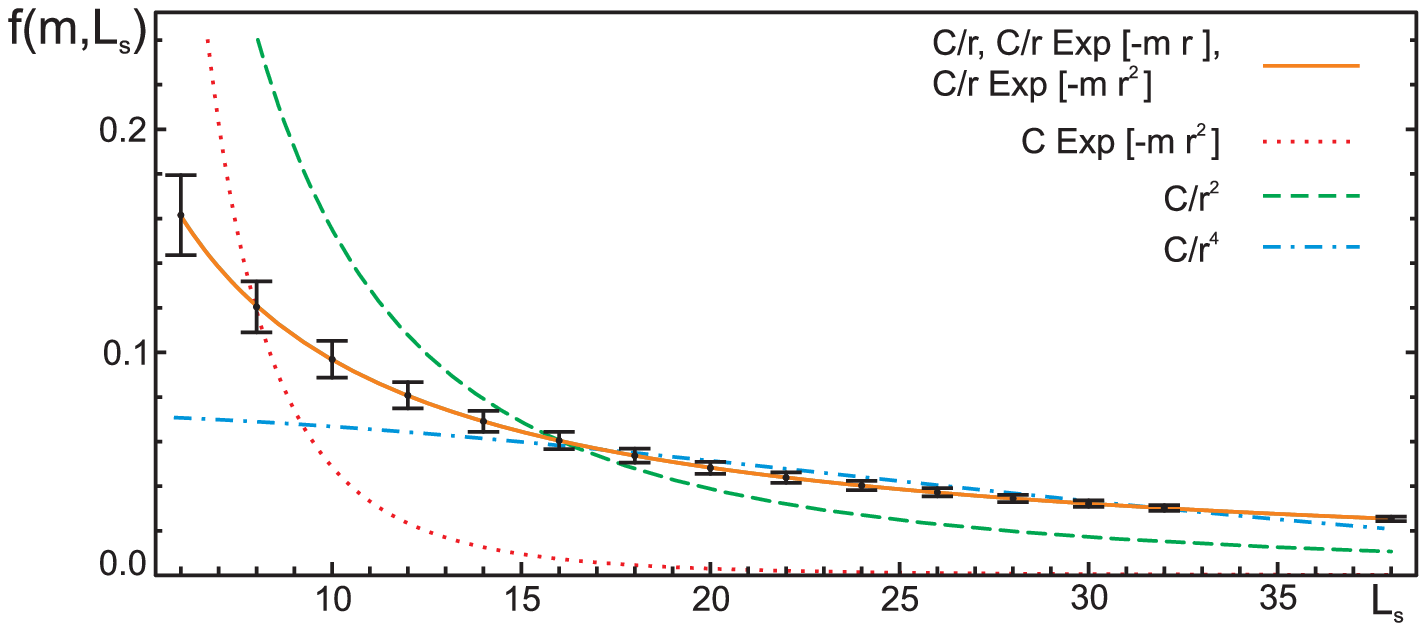}
\end{center}\label{Fig:2}
{{\bf Figure 2:} $f(m, L_s)$ versus $L_s$ and fitting curves ($L_t
= 4$, $\beta = 2.6$).} \vskip 0.5cm

\parbox[b][85mm][t]{30mm}{
\begin{center}
{\scriptsize\vskip 0.52cm
\begin{tabular}[c]{|l|c|}
 \hline\rule{0pt}{1ex}{\small\bf $f(m,L_S)$} & {\bf$L_S$}\\\hline\hline
 \rule{0pt}{0.5ex}\scriptsize$0.161\pm0.018$      &  $6$\\
 \rule{0pt}{0.5ex}$0.12\pm0.011$       &  $8$\\
 \rule{0pt}{0.5ex}$0.097\pm0.008$      &  $10$\\
 \rule{0pt}{0.5ex}$0.081\pm0.006$      &  $12$\\
 \rule{0pt}{0.5ex}$0.069\pm0.005$      &  $14$\\
 \rule{0pt}{0.5ex}$0.06\pm0.004$       &  $16$\\
 \rule{0pt}{0.5ex}$0.054\pm0.003$      &  $18$\\
 \rule{0pt}{0.5ex}$0.048\pm0.003$      &  $20$\\
 \rule{0pt}{0.5ex}$0.044\pm0.002$      &  $22$\\
 \rule{0pt}{0.5ex}$0.04\pm0.002$       &  $24$\\
 \rule{0pt}{0.5ex}$0.037\pm0.002$      &  $26$\\
 \rule{0pt}{0.5ex}$0.0345\pm0.0016$    &  $28$\\
 \rule{0pt}{0.5ex}$0.0322\pm0.0014$    &  $30$\\
 \rule{0pt}{0.5ex}$0.0302\pm0.0013$    &  $32$\\
 \rule{0pt}{0.5ex}$0.025\pm0.001$      &  $38$\\\hline
\end{tabular}
\small
\vskip 0.3cm {\bf{Table 1:}}
\mbox{{{Monte-Carlo data.}}} }
\end{center}
}
\hfill
\parbox[b][85mm][t]{100mm}{
\begin{center}
{\scriptsize \vskip 0.5cm
\begin{tabular}[c]{|l|c|c|c|}
 \cline{2-4} \multicolumn{1}{c}{~} & \multicolumn{3}{|c|}{\rule{0pt}{3ex}\small\bf Abelian field} \\
 \hline\rule{0pt}{3ex} {\bf Fit function} & $\chi^2$ & $C$ & $m$ \\\hline\hline
 \rule{0pt}{3ex}\scriptsize$C\exp(-mr)$           & $901.8$   & $0.063$ & $m = (2.44^{+0.06}_{-0.06})\times10^{-2}$ \\
 \rule{0pt}{3ex}$C\exp(-m^2r^2)$       & $1924.4$  & $0.035$ & $m = (1.57^{+0.02}_{-0.02})\times10^{-2}$ \\\hline
 \rule{0pt}{3ex}$C/r$                  & $7.090$   & $0.911$ &                           \\
 \rule{0pt}{3ex}$C/r\exp(-mr)$         & $7.086$   & $0.912$ & $m = (1.25^{+52}_{-54})\times10^{-6}$ \\
 \rule{0pt}{3ex}$C/r\exp(-m^2r^2)$     & $7.090$   & $0.911$ & $m^2 = (2.4^{+5951.2}_{-5784})\times10^{-10}$ \\\hline
 \rule{0pt}{3ex}$C/{r^2}$              & $31400$   & $28.13$ &                           \\
 \rule{0pt}{3ex}$C/{r^2}\exp(-m^2r^2)$ & $7550$    & $18.26$ & $m^2 = -3.3\times10^{-5} $\\\hline
 \rule{0pt}{3ex}$C/{r^4}$              & $159500$  & $248.9$ &                           \\
 \rule{0pt}{3ex}$C/{r^4}\exp(-m^4r^4)$ & $161000$  & $10.0$  & $m = 0.0$ \\\hline
\end{tabular}
\small\vskip 0.3cm ~\\\noindent{\bf{Table 2:}}
\mbox{\parbox[t]{0.64\hsize}{{Fit results for magnetic mass
of Abelian magnetic field.}}} ~\\~ }
\end{center}
}

The result obtained is unexpected one. In fact, we assumed to find
a nonzero value of the order $m_{magn.} \sim g^2 T$, as at zero
external field \cite{DeGrand:1981yx}. To be sure in our analysis,
we have reproduced  the later result as well. It should also be
noted that due to a huge amount of data we have guarantied that
the absolute value of errors is of $10^{2}-10^3$ times less than
the value of the corresponding quantity.
\section{Discussion}
We performed calculation of the magnetic mass for  neutral gluons
in the Abelian chromomagnetic field at high temperature. Such type
fields have to be spontaneously generated in deconfinement phase.
They are stable due to large value of the charged gluon magnetic
mass \cite{Skalozub:1999bf}, \cite{Demchik:2006qj}. The results
obtained in continuum field theory coincide with that of MC
simulations on a lattice. In both cases zero value is determined
with the accuracy proper to the methods used. Hence, we conclude
that such magnetic fields are long range ones. This, in
particular, means that Abelian magnetic fields, being the
solutions to the non-Abelian gauge field equations without
sources, are spontaneously created at high temperature and exist
till the confinement phase transition happens. This also concerns
the electroweak sector of the standard model. In this case only
the non-Abelian constituent of the magnetic field related to the
$SU(2)$ weak isospin group is spontaneously created  at high
temperature. The constituent related to the weak hypercharge
subgroup $U(1)_Y$ is zero.

Interesting additional arguments in favor of spontaneous vacuum
magnetization at high temperature were obtained in sect. 3. In the
measurements fulfilled,  we observed that for the fitting function
$f(m, L_s) = C/r^2 $ corresponding to the magnetic flux tube
formation  the $\chi^2$ value is very large and entirely
inconsistent with the data. But in the geometry of measurements it
describes the conservation of the magnetic flux introduced by the
twist of the boundary conditions. The best fit functions $C/r$,
$C/r \exp( - m r )$ with  very small (actually, zero) $m$ are
signaling an increase of the mean magnetic field strength
penetrating the plaquette perpendicular to the field direction. As
a result, the flux though  the whole $(X-Y)$ plain should
increase. The only natural explanation  is the spontaneous
generation of the field inside the volume of the lattice.

Since the field created at high temperature is described by the
solution Eq.\Ref{field} which spoils gauge invariance, the
question of its physical content arises. One of the possible ways
to restore gauge invariance consists in formation of the domain
structure having a special boundary which ensures the invariance.
Other possibility consists in spontaneous breaking of color
symmetry at high temperatures having macroscopic consequences and
some remnants at low temperatures after the confinement phase
transition (the electroweak phase transition) happens. These
alternatives as well as other scenarios  require  separate
investigations.

As a general conclusion we  note that the presence of Abelian
temperature dependent magnetic fields in high temperature phase of
QCD and other gauge field theories has to be taken into
consideration when various phenomena are investigated.
\section{Acknowledgements}
One of us (VS) was supported by DFG under Grant No BO1112/16-1. He
also thanks the Institute for Theoretical Physics of Leipzig
University for kind hospitality.



\begin{thebibliography}{10}
\bibitem{Cea:2007yv}
  P.~Cea, L.~Cosmai and M.~D'Elia,
  JHEP {\bf 0712}, 097 (2007).
\bibitem{Starinets:1994vi}
  A.~O.~Starinets, A.~S.~Vshivtsev and V.~C.~Zhukovsky,
  Phys.\ Lett.\  B {\bf 322}, 403 (1994).
\bibitem{Enqvist:1994rm}
  K.~Enqvist and P.~Olesen,
  Phys.\ Lett.\  B {\bf 329}, 195 (1994)
  [arXiv:hep-ph/9402295].
\bibitem{Skalozub:1999bf}
  V.~Skalozub and M.~Bordag,
  Nucl.\ Phys.\  B {\bf 576}, 430 (2000)
  [arXiv:hep-ph/9905302].
\bibitem{Skalozub:1996ax}
  V.~V.~Skalozub,
  Int.\ J.\ Mod.\ Phys.\  A {\bf 11}, 5643 (1996).
\bibitem{Demchik:2008zz}
  V.~I.~Demchik and V.~V.~Skalozub,
  Phys.\ Atom.\ Nucl.\  {\bf 71}, 180 (2008).
\bibitem{Savvidy:1977as}
  G.~K.~Savvidy,
  Phys.\ Lett.\  B {\bf 71}, 133 (1977).
\bibitem{Ebert:1996tj}
  D.~Ebert, V.~C.~Zhukovsky and A.~S.~Vshivtsev,
  Int.\ J.\ Mod.\ Phys.\  A {\bf 13}, 1723 (1998).
\bibitem{Strelchenko:2004 eg}
V.V. Skalozub and A.V. Strelchenko,
Eur. \ Phys.\ J. C {\bf 33}, 105 (2004).
\bibitem{Bordag:2006pr}
  M.~Bordag and V.~Skalozub,
  Phys.\ Rev.\  D {\bf 75}, 125003 (2007)
  [arXiv:hep-th/0611256].
\bibitem{Bordag:2008wp}
  M.~Bordag and V.~Skalozub,
  Phys.\ Rev.\  D {\bf 77}, 105013 (2008)~\\~
  [arXiv:0801.2306 [hep-th]].
\bibitem{Schwinger:1973kp}
  J.~S.~Schwinger,
  Phys.\ Rev.\  D {\bf 7}, 1696 (1973).
\bibitem{Bordag:2005br}
  M.~Bordag and V.~Skalozub,
  Eur.\ Phys.\ J.\  C {\bf 45}, 159 (2006)
  [arXiv:hep-th/0507141].
\bibitem{Tsai:1974id}
  W.~Y.~Tsai and A.~Yildiz,
  Phys.\ Rev.\  D {\bf 8}, 3446 (1973)
  [Erratum-ibid.\  D {\bf 9}, 2489 (1974)].
\bibitem{Kalashnikov:1982sc}
  O.~K.~Kalashnikov,
  Fortsch.\ Phys.\  {\bf 32}, 525 (1984).
\bibitem{Demchik:2009ni}
  V.~Demchik and A.~Strelchenko,
  arXiv:0903.3053 [hep-lat].
  \bibitem{Demchik:2010fd}
  V.~Demchik,
  Comp.\ Phys.\ Comm. {\bf 182}, 692 (2011), \\ doi: 10.1016/j.cpc.2010.12.008;
  arXiv:1003.1898 [hep-lat].
\bibitem{Demchik:2006qj}
  V.~Demchik and V.~Skalozub,
  arXiv:hep-lat/0601035.
\bibitem{DeGrand:1981yx}
  T.~A.~DeGrand and D.~Toussaint,
  Phys.\ Rev.\  D {\bf 25} (1982) 526.
    Phys.\ Rev.\  D {\bf 25} (1982) 526.
\bibitem{'tHooft:1979uj}
  G.~'t Hooft,
  Nucl.\ Phys.\  B {\bf 153}, 141 (1979).
\bibitem{Creutz:1987xi}
  M.~Creutz,
  Phys.\ Rev.\  D {\bf 36} (1987) 515.
\end{thebibliography}
\end{document}